\documentclass[twocolumn,prl,aps,superscriptaddress]{revtex4}

\usepackage{amsthm}
\usepackage{amsmath,latexsym,amssymb,verbatim,enumerate,graphicx}
\usepackage{hyperref}
\usepackage{color}

\newtheorem{thm}{Theorem}
\newtheorem*{thm*}{Theorem}

\newtheorem*{prop*}{Proposition}
\newtheorem{lem}[thm]{Lemma}
\newtheorem*{lem*}{Lemma}

\newtheorem*{fact*}{Fact}
\newtheorem{cor}[thm]{Corollary}
\newtheorem*{cor*}{Corollary}

\usepackage{prettyref}
\newcommand{\savehyperref}[2]{\texorpdfstring{\hyperref[#1]{#2}}{#2}}

\newrefformat{eq}{\savehyperref{#1}{\textup{(\ref*{#1})}}}
\newrefformat{lem}{\savehyperref{#1}{Lemma~\ref*{#1}}}
\newrefformat{def}{\savehyperref{#1}{Definition~\ref*{#1}}}
\newrefformat{thm}{\savehyperref{#1}{Theorem~\ref*{#1}}}
\newrefformat{cor}{\savehyperref{#1}{Corollary~\ref*{#1}}}
\newrefformat{cha}{\savehyperref{#1}{Chapter~\ref*{#1}}}
\newrefformat{sec}{\savehyperref{#1}{Section~\ref*{#1}}}
\newrefformat{app}{\savehyperref{#1}{Appendix~\ref*{#1}}}
\newrefformat{tab}{\savehyperref{#1}{Table~\ref*{#1}}}
\newrefformat{fig}{\savehyperref{#1}{Figure~\ref*{#1}}}
\newrefformat{hyp}{\savehyperref{#1}{Hypothesis~\ref*{#1}}}
\newrefformat{alg}{\savehyperref{#1}{Algorithm~\ref*{#1}}}
\newrefformat{rem}{\savehyperref{#1}{Remark~\ref*{#1}}}
\newrefformat{item}{\savehyperref{#1}{Item~\ref*{#1}}}
\newrefformat{step}{\savehyperref{#1}{step~\ref*{#1}}}
\newrefformat{conj}{\savehyperref{#1}{Conjecture~\ref*{#1}}}
\newrefformat{fact}{\savehyperref{#1}{Fact~\ref*{#1}}}
\newrefformat{prop}{\savehyperref{#1}{Proposition~\ref*{#1}}}
\newrefformat{prob}{\savehyperref{#1}{Problem~\ref*{#1}}}
\newrefformat{claim}{\savehyperref{#1}{Claim~\ref*{#1}}}
\newrefformat{relax}{\savehyperref{#1}{Relaxation~\ref*{#1}}}
\newrefformat{red}{\savehyperref{#1}{Reduction~\ref*{#1}}}
\newrefformat{part}{\savehyperref{#1}{Part~\ref*{#1}}}

\newcommand{\MYstore}[2]{%
  \global\expandafter \def \csname MYMEMORY #1 \endcsname{#2}%
}

\newcommand{\MYload}[1]{%
  \csname MYMEMORY #1 \endcsname%
}

\newcommand{\MYnewlabel}[1]{%
  \newcommand\MYcurrentlabel{#1}%
  \MYoldlabel{#1}%
}

\newcommand{\MYdummylabel}[1]{}

\newcommand{\torestate}[1]{%
  \let\MYoldlabel\label%
  \let\label\MYnewlabel%
  #1%
  \MYstore{\MYcurrentlabel}{#1}%
  \let\label\MYoldlabel%
}

\newcommand{\restatethm}[1]{%
  \let\MYoldlabel\label
  \let\label\MYdummylabel
  \begin{thm*}[Restatement of \prettyref{#1}]
    \MYload{#1}
  \end{thm*}
  \let\label\MYoldlabel
}

\newcommand{\restatelem}[1]{%
  \let\MYoldlabel\label
  \let\label\MYdummylabel
  \begin{lem*}[Restatement of \prettyref{#1}]
    \MYload{#1}
  \end{lem*}
  \let\label\MYoldlabel
}

\newcommand{\restatecor}[1]{%
  \let\MYoldlabel\label
  \let\label\MYdummylabel
  \begin{cor*}[Restatement of \prettyref{#1}]
    \MYload{#1}
  \end{cor*}
  \let\label\MYoldlabel
}

\newcommand{\restateprop}[1]{%
  \let\MYoldlabel\label
  \let\label\MYdummylabel
  \begin{prop*}[Restatement of \prettyref{#1}]
    \MYload{#1}
  \end{prop*}
  \let\label\MYoldlabel
}

\newcommand{\restatefact}[1]{%
  \let\MYoldlabel\label
  \let\label\MYdummylabel
  \begin{fact*}[Restatement of \prettyref{#1}]
    \MYload{#1}
  \end{fact*}
  \let\label\MYoldlabel
}

\newcommand{\restate}[1]{%
  \let\MYoldlabel\label
  \let\label\MYdummylabel
  \MYload{#1}
  \let\label\MYoldlabel
}

\newtheorem*{rep@theorem}{\rep@title}
\newcommand{\newreptheorem}[2]{%
\newenvironment{rep#1}[1]{%
 \def\rep@title{#2 \ref{##1} (restatement)}%
 \begin{rep@theorem}}%
 {\end{rep@theorem}}}
\makeatother

\newreptheorem{thm}{Theorem}
\newreptheorem{lem}{Lemma}
\newreptheorem{cor}{Corollary}

\def\ba#1\ea{\begin{align}#1\end{align}}
\def\ban#1\ean{\begin{align*}#1\end{align*}}

\newcommand{\ot}{\otimes}
\newcommand{\be}{\begin{equation}}
\newcommand{\ee}{\end{equation}}

\def\benum{\begin{enumerate}}
\def\eenum{\end{enumerate}}

\def\ghaar{G_{\text{Haar}}}

\def\squareforqed{\hbox{\rlap{$\sqcap$}$\sqcup$}}
\def\qed{\ifmmode\squareforqed\else{\unskip\nobreak\hfil
\penalty50\hskip1em\null\nobreak\hfil\squareforqed
\parfillskip=0pt\finalhyphendemerits=0\endgraf}\fi}
\def\endenv{\ifmmode\;\else{\unskip\nobreak\hfil
\penalty50\hskip1em\null\nobreak\hfil\;
\parfillskip=0pt\finalhyphendemerits=0\endgraf}\fi}

\newcommand{\bra}[1]{\langle #1|}
\newcommand{\ket}[1]{|#1\rangle}

\newcommand{\ben}{\begin{equation}}
\newcommand{\een}{\end{equation}}

\newcommand{\<}{\langle}
\renewcommand{\>}{\rangle}

\def\L{\left}
\def\R{\right}

\DeclareMathOperator{\tr}{tr}
\def\be{\begin{equation}}
\def\ee{\end{equation}}
\def\ben{\begin{eqnarray}}
\def\een{\end{eqnarray}}
\def\ot{\otimes}

\def\bei{\begin{itemize}}
\def\eei{\end{itemize}}

\def\eps{\epsilon}

\def\cS{{\cal S}}

\mathchardef\ordinarycolon\mathcode`\:
\mathcode`\:=\string"8000
\def\vcentcolon{\mathrel{\mathop\ordinarycolon}}
\begingroup \catcode`\:=\active
  \lowercase{\endgroup
  \let :\vcentcolon
  }

\newcommand{\nc}{\newcommand}
\nc{\rnc}{\renewcommand} \nc{\beq}{\begin{equation}}
\nc{\eeq}{{\end{equation}}} \nc{\bea}{\begin{eqnarray}}
\nc{\eea}{\end{eqnarray}} \nc{\beqa}{\begin{eqnarray}}
\nc{\eeqa}{\end{eqnarray}} \nc{\lbar}[1]{\overline{#1}}
 \nc{\proj}[1]{|#1\rangle\!\langle #1 |} 
\nc{\avg}[1]{\langle#1\rangle}

\nc{\conv}{\operatorname{conv}}
\nc{\smfrac}[2]{\mbox{$\frac{#1}{#2}$}} \nc{\Tr}{\operatorname{Tr}}
\nc{\ox}{\otimes} \nc{\dg}{\dagger} \nc{\dn}{\downarrow}
\nc{\lmax}{\lambda_{\text{max}}}
\nc{\lmin}{\lambda_{\text{min}}}

\nc{\csupp}{{\operatorname{csupp}}}
\nc{\qsupp}{{\operatorname{qsupp}}} \nc{\var}{\operatorname{var}}
\nc{\rar}{\rightarrow} \nc{\lrar}{\longrightarrow}
\nc{\poly}{\operatorname{poly}}
\nc{\polylog}{\operatorname{polylog}} \nc{\Lip}{\operatorname{Lip}}
\nc{\Om}{\Omega}
\nc{\wt}[1]{\widetilde{#1}}

\def\>{\rangle}
\def\<{\langle}

\nc{\glneq}{{\raisebox{0.6ex}{$>$}  \hspace*{-1.8ex} \raisebox{-0.6ex}{$<$}}}
\nc{\gleq}{{\raisebox{0.6ex}{$\geq$}\hspace*{-1.8ex} \raisebox{-0.6ex}{$\leq$}}}

\nc{\vholder}[1]{\rule{0pt}{#1}}
\nc{\wh}[1]{\widehat{#1}}
\nc{\h}[1]{\widehat{#1}}

\nc{\ob}[1]{#1}

\def\beq{\begin {equation}}
\def\eeq{\end {equation}}

\def\be{\begin{equation}}
\def\ee{\end{equation}}

\nc{\eq}[1]{(\ref{eq:#1})} 
\nc{\eqs}[2]{\eq{#1} and \eq{#2}}

\nc{\eqn}[1]{Eq.~(\ref{eqn:#1})}
\nc{\eqns}[2]{Eqs.~(\ref{eqn:#1}) and (\ref{eqn:#2})}
\newcommand{\fig}[1]{Fig.~\ref{fig:#1}}

\newcommand{\thmref}[1]{Theorem~\ref{thm:#1}}

\newcommand{\corref}[1]{Corollary~\ref{cor:#1}}

\nc{\region}{\cS\cW}

\begin{document}

\title{Efficient Quantum Pseudorandomness}

\author{Fernando G.S.L. Brand\~ao}
\affiliation{Microsoft Research Redmond}
\affiliation{Computer Science Department, University College London}
\author{Aram W. Harrow}
\affiliation{Center for Theoretical Physics, Massachusetts Institute of Technology}
\author{Micha\l{} Horodecki}
\affiliation{National Quantum Information Center of Gdansk}

\begin{abstract}
Randomness is both a useful way to model natural systems and a useful tool for engineered systems, e.g. in computation, communication and control.  Fully random transformations require exponential time for either classical or quantum systems, but in many case pseudorandom operations can emulate certain properties of truly random ones. Indeed in the classical realm there is by now a well-developed theory of such pseudorandom operations. However the construction of such objects turns out to be much harder in the quantum case. Here we show that random quantum circuits are a powerful source of quantum pseudorandomness. This gives the for the first time a polynomial-time construction of quantum unitary designs, which can replace fully random operations in most applications, and shows that generic quantum dynamics cannot be distinguished from truly random processes. We discuss applications of our result to quantum information science, cryptography and to understanding self-equilibration of closed quantum dynamics. 

\end{abstract}
\maketitle

Random processes are ubiquitous both in natural and engineered
systems. They are both an effective way to model many systems, and
also a vital tool in algorithms, communication, control, cryptography,
and elsewhere. However a random function on $n$ bits is known to
require $\exp(\Omega(n))$ elementary operations\footnote{ $\Omega(f(n))$ refers to a
  function that is $\geq c f(n)$ for some $c>0$ and for sufficiently
  large $n$.} to implement and a similar number of random bits to even
specify~\cite{Shannon49}, meaning that in fact such random functions
are neither found in nature, nor can they can be designed on a
computer. Instead, we now know many methods for engineering {\em
  pseudorandom} functions using far less randomness. These
pseudorandom functions can be proven to be indistinguishable from
truly random functions either by any test that examines their first
$k$ moments \cite{ABI86} (in which case they are called $k$-designs),
or by any computationally limited test \cite{GGM86} (for which case
the term {\em pseudorandom function} is usually reserved).  This
Letter will focus on $k$-designs.

While these constructions mean that a carefully designed algorithm can simulate a random function in many circumstances, they do not speak to the question of whether we should expect natural processes to also resemble random functions. However, it was proved by \cite{HooryBrodsky04, BH08} that even reasonably short random reversible circuits yield approximate $k$-designs, meaning that they approximate well the first $k$ moments of a truly random function. These circuits are defined to be sequences of basic reversible operations, each involving three bits, which is the simplest type of reversible circuit that is computationally universal.  As such, they form plausible toy models for the dynamics of actual systems, and provide some rigorous justification for the intuition that generic dynamics cannot easily be distinguished from fully random functions.

In recent decades, quantum mechanics has been found to dramatically change the nature of information and information-processing~\cite{Nielsen-Chuang}, implying among other things potential new applications such as quantum cryptography and computation.  The above story needs then to be 
modified to account for quantum mechanics. The problem of quantum pseudorandomness was posed in \cite{Emerson03} where it was asked to what extent random quantum circuits can mimic random quantum transformations.
The simplest quantum systems are 2-level systems, called {\em qubits}, and any larger quantum system can be decomposed into some number $n$ of qubits (e.g. the state of $m$ fermions in $n$ modes can be expressed using $\lceil \binom{n}{m}\rceil$ qubits).  A quantum circuit is a sequence of gates, each acting on a constant number of qubits.  Short quantum circuits are roughly equivalent in power to time-evolution by local Hamiltonians for short times.
The uniform distribution over unitary matrices is called the {\em Haar measure} and, as in the classical case, has been extensively studied as a model of natural generic dynamics \cite{Haa91} with applications to black holes \cite{HP07,Sus14}, quantum information processing \cite{ADHW06, HLSW04, Sen06} and elsewhere. In a further parallel to the classical case, Haar-random unitary matrices on $n$ qubits cannot be implemented, even approximately, without $\Omega(4^n)$ elementary operations and $\Omega(4^n)$ random bits~\cite{Knill95}. 

We thus have the same need for quantum pseudorandomness and an
analogous notion of unitary $k$-designs. Again unitary $k$-designs can
be used in place of Haar-random unitaries in most applications
(e.g. for encoding quantum information to protect from errors
\cite{ADHW06}, or realizing more efficient quantum process tomography
\cite{Emerson03}), but it has been much harder to prove that efficient
unitary $k$-designs exist. It was long conjectured that random quantum
circuits yield approximate $k$-designs, but for a long time, this was
known only for $k=1$; in other words, only the first moments of a quantum
state were known to rapidly equilibrate under random dynamics.
However, 1-designs can be realized even without entangling operations
(e.g. a random product of Pauli matrices will suffice), while Haar-uniform
unitaries create states with large amounts of
entanglement~\cite{HLW06}, so 1-designs do not give a qualitatively
good fit for the Haar measure.  A better, but still imperfect, goal is
to achieve a 2-design.  It has been known how to efficiently engineer
a 2-design on a quantum computer by selecting a random element of the
so-called {\it Clifford group}---a restricted class of quantum
gates---which involves creating significant entanglement while still
performing operations far simpler than those resulting from the Haar
measure~\cite{DLT02}.  

Initial numerical work suggested that indeed random quantum circuits
were approximate 2-designs~\cite{Emerson03, AB08}. Later work was able to establish that random circuits
matched the entanglement properties of 2-designs \cite{ODP06, DOP07},
and finally that they in fact were approximate 2-designs \cite{HL09,
  DJ10, AB08, Zni08}.  Since even the Clifford group, which is not
universal for quantum computation, yields a 2-design, this too is a
limited proxy for the Haar measure.  Later work achieved 3-designs
\cite{HL09, BH10} (see also \cite{CHMPS13}). In this work we settle the question and achieve
$k$-designs for any $k$ via circuits of length $\poly(n,k)$.  Full
details are given in \cite{BHH12}, where it is also shown that this
work cannot be substantially improved.  We follow part of the
framework of \cite{BV10}, which conjectured our result and gave a
mean-field argument supporting it.

\paragraph{Definitions:} Let us give a more precise definition of
approximate unitary designs. First we say a probability measure $\mu$
on $\mathbb{U}(d)$ (the group of $d \times d$ unitary matrices) is a
unitary $k$-design if for every monomial $q(U) = U_{i_1 j_1} \ldots
U_{i_k j_k} U^{*}_{m_1 n_1} \ldots U^{*}_{m_k n_k}$ of degree at most
$k$, in the entries of the unitary $U_{ij}$ and their complex
conjugate $U^{*}_{nm}$, the average of $q(U)$ over the Haar measure is
the same as the average over $\mu$.  

In turn, we say a measure $\mu$ on $\mathbb{U}(d)$ forms an $\varepsilon$-approximate $k$-design if 
\begin{equation} \label{eq:approxdesign}
\left |  \mathbb{E}_{\text{Haar}} q(U)  -   \mathbb{E}_{\mu} q(U)   \right | \leq \varepsilon
\end{equation}
with $\mathbb{E}_{\text{Haar}}, \mathbb{E}_{\mu}$ the expectations over the Haar measure and $\mu$, respectively. There are other definitions of $\varepsilon$-approximate $k$-designs \cite{BHH12}, but they turn out to be equivalent to the one above, up to a rescale of the approximation factor~\cite{Low10}.

\begin{figure}
\begin{center}  
\includegraphics[width=0.9\columnwidth]{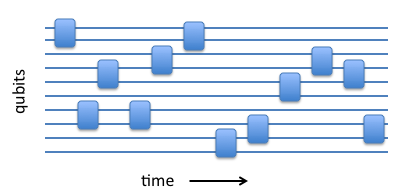}
\caption{In our model of a random quantum circuit, there are $n$
  qubits on a line, here arranged vertically.  Time proceeds from left
  to right.  In each time step, two adjacent qubits are chosen at
  random and a random two-qubit unitary operator is applied to them.
 \label{fig:circuit}}  
\end{center}  
\end{figure}

We model random quantum circuits as
random walks on $\mathbb{U}(2^{n})$ following Refs.~\cite{HP07,BH10}.
In each step of the walk an index $i$ is chosen uniformly at random from $\{1,\ldots,n-1\}$ and a unitary $U_{i,i+1}$ drawn from the Haar measure on $\mathbb{U}(4)$ is applied to the two neighboring qubits $i$ and $i+1$.   This is illustrated in
\fig{circuit}.
Other choices of random circuits are possible (e.g. considering non-nearest-neighbor gates or gates from different universal sets as in Ref.~\cite{EWS03}) and variants of our results will apply to them as well.  However here for concreteness we focus on the model above.  A related model of random walk on the unitary (or orthogonal) group is Kac's random walk, extensively studied in connection to statistical mechanics (see e.g. \cite{Oli07}). 


Our main result is the following:

\begin{thm}\label{thm:main-design} \mbox{}
Random circuits with $O(nk^{9}(nk+ \log(1/\eps)))$ gates
form $\eps$-approximate $k$-designs. 
\end{thm}


\paragraph{Applications:} As $k$-wise independent distributions
(or classical $k$-designs) have widespread applications, so too do
approximate unitary $k$-designs \cite{ADHW06, Emerson03,Low09}. 

Here we briefly outline one application of our result to the problem of understanding equilibration in closed quantum dynamics. Consider the unitary time evolution of a system, initially in a fixed state, say all spins up $\ket{\uparrow}^{\otimes n}$. The total state at any particular time is pure and hence does not appear to equilibrate in any sense. However a long sequence of investigations, starting with von Neumann in 1929 \cite{Neu29}, has elucidated that the state does equilibrate if one imposes constraints on the kind of observations possible \cite{LPSW09, GLTZ06, TCFM12, BCH11}. For instance suppose that one only has access to measurements on a few of the particles. Then it turns out that the building up of entanglement in the quantum state leads to local equilibration of every small subset of particles, for almost all times \cite{LPSW09}. The limits of equilibration in closed quantum dynamics is an interesting problem. What is the largest class of observables for which equilibration holds? Our result on unitary designs allows us advance this question significantly. 

Define the circuit complexity of a measurement  as the minimum size of any circuit of two-qubit gates that implements the measurement. Physical measurements (e.g. measurement of magnetization or heat capacity or even topological invariants) generally have low complexity. We show that in generic quantum dynamics given by random circuits (which model the case of generic evolutions under time-dependent Hamiltonians), the system equilibrates with respect to all measurements of low complexity.  A general quantum measurement with two outcomes can be represented by operators $M$ and $I-M$, with $0\leq M\leq I$. We then have:

\begin{cor} \label{cor:fool-M}
For every $k \geq 1$, for sufficiently large $n$ and almost all random
circuits $U$ of size $O(n^{11k+9})$ on $n$ qubits,  
\begin{equation}
\left|  \bra{\uparrow^{\otimes n}} U^\dag MU \ket{\uparrow^{\otimes n}} -  \frac{\tr(M)}{2^n}    \right | \leq 2^{- n/4}.  
\label{eq:fool-M}\end{equation}
for every measurement $\{M, I - M \}$ of circuit complexity less than $n^k$.
\end{cor}

{\em Proof sketch:}  If $M$ is fixed and $U$ is Haar uniform then
\eq{fool-M} holds with high probability; large deviations are
suppressed with probability exponential in the dimension, meaning
$\exp(-2^{\Omega(n)})$.  If instead $U$ is drawn from 
a $t$-design then this probability becomes $\exp(-t)$.  We can
approximate any low-complexity $M$ with a measurement drawn from a
set of size roughly $\exp(n^k)$.  Thus \eq{fool-M} holds
(approximately) for {\em all} low-complexity $M$ with probability
$\leq \exp(n^k - t)$.  For this to be $\ll 1$ we need $t \gg n^k$,
which according to \thmref{main-design} can be achieved by a random
circuit of length $O(n^{11k + 9})$.  The full details of this proof are in
\cite{BHH12}.


Another interpretation of \corref{fool-M} is in the context of quantum
cryptography. It gives a procedure for so called {\it quantum data
  hiding} against a computationally-bounded adversary, meaning that
information is present in a state but cannot be measured without using
a long quantum computation (cf.~\cite{DLT02}). Indeed, Corollary
\ref{cor:fool-M} gives that all but a $2^{-\Omega(n)}$-fraction of
states generated by circuits of size $O(n^{11k+9})$ cannot be
distinguished from the maximally mixed state with bias larger than
$n^{- \Omega(1)}$ by any circuit of size $n^k$. So whether one has the
particular pure state or the maximally mixed state is hidden from any
adversary that is constrained to run in time $n^k$.  

One situation where the assumption is satisfied is when the adversary has less computational power
than the honest parties. Although this might be the case sometimes, it is admittedly not a very realistic
assumption. Another situation, perhaps more relevant, is when the time it takes the adversary to
decode the message is longer than the time is takes to send the message from one honest party to the other. 
In this case the honest parties can abort the protocol if the message is not delivered in time.

\paragraph{Proof overview:} The remainder of the paper will
give a high-level description of the proof of Theorem
\ref{thm:main-design}.  A full proof is in \cite{BHH12}.
The proof is based on an interplay of
techniques from quantum many-body theory~\cite{Nac96}, representation theory~\cite{approx-ortho}, and
the theory of Markov chains~\cite{Oli07}, and we believe similar ideas might find
further applications elsewhere.  

{\it Expressing the problem in terms of spectral properties of matrices: Classical case.}  As a warmup to understanding the properties of our random circuits, consider the classical analogue.  If $C$ is a $t$-gate reversible classical circuit acting on $n$ bits, then we can think of it as a permutation matrix of size $2^n$.  Since mixing over the set of all permutations requires exponentially long circuits, we instead examine the behavior of the moments of the circuit.  To represent the circuit's $k$'th moments, we can examine its action on sets of $k$ inputs, each of which are $n$-bit strings.  Using the tensor product, this action can be also be described as a matrix: $C^{\ot k}$, which maps $\ket{i_1}\ot \ldots\ot \ket{i_k}$ to $C\ket{i_1} \ot \cdots \ot C\ket{i_k}$.  The advantage of this representation is that the average over $t$-step circuits of $C^{\ot
  k}$ (call it $A_{t,k}$) is simply the $t^{\text{th}}$ power of the
average over one-step circuits $A_{1,k}$; i.e. $A_{t,k} = A_{1,k}^t$.
Moreover, if the gate set is universal then $A_{t,k}$ will approach
the average over all permutations as $t\rar\infty$.


Determining the rate of convergence now reduces to an eigenvalue problem. Since $A_{\infty,k}=A_{1,k}^\infty$, it must have only eigenvalues 0 or 1.  The 1-eigenspace corresponds to the degrees of freedom that are preserved when the same circuit is applied to each of $i_1,\ldots,i_k$; e.g. information about whether $i_1=i_2$ or $i_1\neq i_2$.  When the set of gates is universal, the matrix $A_{1,k}$ has the same eigenspace with eigenvalue one and has all other eigenvalues smaller than one (i.e. there are no additional "constants of motion").  Thus everything orthogonal to this subspace will decay to 0 as $t\rar\infty$ at a rate controlled by the eigenvalues of $A_{1,k}$.

Our distance after $t$ steps to the average over random permutations 
can be quantified by $\|A_{t,k} - A_{\infty,k}\|$.   Due to the above arguments, this is given just by $(1-\delta)^t$,
where $1-\delta$ is the second-largest eigenvalue of $A_{1,k}$ (disregarding multiplicity). This is the source of the exponential convergence
typically exhibited by Markov chains on discrete state spaces.  As a result, error $\epsilon$ is achieved by taking circuit length $t\geq \delta^{-1}  \log\frac{1}{\epsilon}$.  By proving~\cite{HooryBrodsky04, BH08} that $\delta \geq 1/\poly(k,n)$, it follows that $n$-bit circuits of length $\poly(k,n)$ have $k$-th moments that approximate those of random permutations.


{\it Expressing the problem in terms of spectral properties of matrices: Quantum 
case.}
In the quantum case, the picture is similar~\cite{BV10} if we replace the action on $n$-bit strings with the 
action on $2^n$-dimensional density matrices.  The $k^{\text{th}}$
moments of this action can be expressed~\cite{HL08,BV10} in terms of the matrix
\be
G_\mu= \int_{\mathbb{U}(2^n)} U^{\otimes k}\otimes (U^{*})^{\otimes k} \mu({\rm d}U).
\label{eq:G}
\ee
where $\mu$ is a distribution over unitary transformations of $n$
qubits.   This matrix can also be thought of as the matrix form of
the map that sends $\rho$ to $\int U^{\ot k}\rho (U^\dag)^{\ot k} \mu({\rm d}U).$ 
If  $\mu$ is taken to be Haar measure, we obtain an analogue of $A_{\infty,k}$. 
To obtain an analogue of $A_{1,k}$ one sets $\mu=\nu\equiv \nu_n$
with $\nu_n$ representing an average over $n-1$ choices of pairs of
neighboring qubits and over a random choice of two-qubit gate applied
to those qubits.   As in the classical case, $\ghaar$ is the projector
onto the 1-eigenspace of $G_\nu$, which we will argue below has
dimension $k!$.  Let $1-\delta$ denote the next largest eigenvalue of
$G_\nu$.  Then we again have that 
\be
\|(G_\nu)^t  - \ghaar\| = (1-\delta)^t, 
\label{eq:lambda_G}
\ee
so the length of the circuit ensuring  $\|(G_\nu)^t  -\ghaar \| \leq \epsilon$ 
is given by 
\be
t=\delta(n,k)^{-1} \log\frac{1}{\epsilon},
\label{eq:t-delta}
\ee  
where we have made explicit the dependence of $\delta$ on $n,k$.
Now, our main result is the following estimate 
\be
\delta(n,k) \geq \Omega\L(\frac{1}{nk^{8.1}\log^2(k)}\R),
\label{eq:main}
\ee
which implies that random circuit of length
$t=O(nk^{8.1}\log^2(k)(nk\log(d)+\log(1/\eps)))$ approximates up to $\epsilon$  
the $k^{\text{th}}$ moment of random unitary, thus proving Theorem \ref{thm:main-design}. 


{\it Connection to many-body theory.}  The matrix $G_{\nu}$ can be expressed in terms of a {\it local Hamiltonian}, bringing the problem within the scope of quantum many-body theory.   The quantity $\delta(n,k)$ that determines circuit length in our case was shown in \cite{BV10, BH10, Zni08} to be directly related to the spectral gap of some Hamiltonian consisting of nearest-neighbor interactions between $n$ $D=4^k$-dimensional systems on a line.
This Hamiltonian does not correspond to a physical system, but tools from many-body theory still apply and can help evaluate the gap.

To construct the Hamiltonian,
note that  our random circuit $\nu$ consists of picking with probability $\frac{1}{n-1}$ a random gate on two adjacent qubits.  Thus 
$G_\nu= \frac1{n-1} \sum_{i=1}^{n-1}P_{i,i+1}$
where  $P_{i, i+1} := \int_{\mathbb{U}(4)} \left(U_{i, i+1}\right)^{\otimes k, k} \mu_{\text{Haar}}({\rm d}U)$
and $U_{i,i+1}$ acts on the $i$-th and $(i+1)$-th qubit.
Define
\begin{equation} \label{Hntdef}
H_{n, k} := (n-1)(I - G_\nu) = \sum_{i = 1}^{n-1} h_{i, i+1}
\end{equation}
with local terms $h_{i, i+1} := I - P_{i, i+1}$ where
$I$ is the identity operator.   It can then be shown that the ``energy'' of a ground state of the Hamiltonian
is zero (corresponding to the 1-eigenspace of $G_\nu$), while that of
the first excited level is  $(n-1)\delta(n,k)$.
In other words, the spectral gap of $H_{n,k}$ (denoted $\Delta(H_{n,k})$) is directly related to the difference of
the largest and second largest eigenvalue of $G_\nu$ according to
$\delta(n,k) = \frac{\Delta(H_{n, k})}{n-1}.$
Thus to determine the length of the random circuit, that approximates $k$-design, 
it sufficies to lower bound the spectral gap of $H_{n, k}$.
\vspace{0.2 cm}

{\it Bounding the Spectral Gap.}
Despite of decades of research on many-body systems, the evaluation of spectral gaps of local Hamiltonians 
is often a formidable task. 
Fortunately, our Hamiltonian has a nice feature of being {\it non-frustrated}: ground states of the total Hamiltonian 
are at the same time a ground states of its local constituents. For such Hamiltonians, Nachtergaele~\cite{Nac96} provided a sufficient condition for existence of a constant gap (in the number of qubits $n$), together with an estimate for the gap.  Nachtergaele's criterion is given in terms of  ground subspaces of Hamiltonians consisting of various numbers of systems $m\leq n$.
One finds that the ground space of our Hamiltonian is spanned by the $k!$ product vectors $\ket{\psi_\sigma}^{\ot n}$, 
labeled by $k$-element permutations $\sigma$. This originates from the fact that the only operators which commute with 
$U^{\ot k}$, where $U$ is arbitrary unitary transformation, are linear combinations of operators that permute systems
(the vectors are actually isomorphic to those operators).  
Were the $\ket{\psi_\sigma}^{\ot n}$ strictly orthogonal, the
Nachtergaele criterion would apply immediately, but this does not hold
here. Yet, by use of group representation theory, we obtain that the vectors
$\ket{\psi_\sigma}^{\ot n}$ have
sufficiently small overlaps (see the Supplementary Materials),
to enable us to apply the criterion and obtain a tight gap.  

As a result, we obtain that the spectral gap of Hamiltonian over $n$
qubits for any $n$ can be bounded by the  gap of Hamiltonian
of a {\it fixed} number of qubits, depending only on the moment $k$:

\begin{lem} \label{LemmaNachtergaele_1}
For every integers $n, k$ with $n \geq  \lceil 2.5\log(4k)\rceil$,
\begin{equation}
\Delta(H_{n, k}) \geq \frac{\Delta(H_{\lceil 2.5 \log (4k) \rceil,
    k})}{4\lceil 2.5 \log (4k)\rceil }.
\label{eq:Hnk_Hk}
\end{equation}
where $\lceil x \rceil$ denotes the least integer no smaller than $x$.
\end{lem}

It remains only to control the gap of the Hamiltonian for systems with
$O(\log(k))$ qubits. Here, since we are concerned with a relatively
small number of qubits, it suffices to establish that  random circuits
on $m$ qubits mix after a number of steps that is exponential in
$m$. We prove this in \cite{BHH12} using the path-coupling method of
\cite{Oli07}.  If $m=O(\log(k))$, then this means a number of steps
that is polynomial in $k$. Translating these mixing bounds back into a
statement about the gap of $H$, we find that $\Delta(H_{O(\log
  k), k}) \geq 1/\poly(k)$.  This completes the proof of
\thmref{main-design}. 

\noindent
{\bf Discussion.} Our main result shows that $O(n^2k^{10})$ random
nearest-neighbor unitary interactions yield a distribution over
unitaries that approximately matches the first $k$ moments of the Haar
measure.   The dependence on $n$ is approximately optimal for 1-d;
indeed, we can think of it (for fixed $k$)  as $O(n)$ rounds of
parallel nearest-neighbor interactions.  Since it takes at least this
much time for information to propagate along a line of $n$ qubits,
this $n$-dependence cannot be improved.  On the other hand, we did not
try hard to reduce the degree of the polynomial in $k$ and it may even
be that the number of gates required is independent of $k$, as in
\cite{BG11}.  Another open question is whether the $n$-dependence
could be improved for better connected geometries;
e.g. nearest-neighbor gates in higher-dimensional lattices or gates
with arbitrary connectivity. 

\thmref{main-design} shows that $k$-designs can be implemented
efficiently on a quantum computer and also describes a plausible
natural process that can give rise to them.  Other work has studied open systems with nonunitary
dynamics~\cite{open-systems} and closed systems with Hamiltonian
dynamics~\cite{MBL15}.  Our model differs from both of these, but
could plausibly arise from a randomly time-varying Hamiltonian.
Still, a major question left open is to understand the conditions under
which realistic physical systems thermalize.

\noindent
{\bf Acknowledgements.}
We would like to thank Dorit Aharonov, Itai Arad, Winton Brown, Daniel
Jonathan, Bruno Nachtergaele, Alex Russell, Tomasz Szarek, and Andreas
Winter. FGSLB acknowledges support from the Swiss National Science
Foundation, via the National Centre of Competence in Research QSIT,
and the Ministry of Education and the National Research Foundation,
Singapore. AWH was funded by NSF grants CCF-1111382 and CCF-1452616 and ARO contract
W911NF-12-1-0486.  MH is supported by EU grant QESSENCE, by Polish
Ministry of Science and Higher Education grant N N202 231937. MH also
acknowledges grant QUASAR of the National Centre for Research and
Development of Poland for preparing the final version of the paper.
Part of this work was done in the National Quantum Information Center
of Gdansk. We thank the hospitality of Institute Mittag Leffler within
the program ``Quantum Information Science,'' where (another) part of
this work was done.

\bibliographystyle{hyperabbrv}
\bibliography{ref}

\end{document}